\documentclass[prb,twocolumn,showpacs,preprintnumbers,amsmath,amssymb]{revtex4}
\usepackage{graphicx}
\usepackage{dcolumn}

\begin{document}

\title{ Synchronous relaxation algorithm for parallel kinetic Monte Carlo}

\author{Yunsic Shim}
\email{yshim@physics.utoledo.edu}
\author{Jacques G. Amar}
\email{jamar@physics.utoledo.edu}
\affiliation{Department of Physics \& Astronomy \\
University of Toledo,  Toledo, OH 43606}

\date{\today}

\begin{abstract}
We investigate the applicability  of the synchronous relaxation (SR) algorithm to  parallel kinetic Monte Carlo simulations of simple models of thin-film growth. A variety of    techniques for optimizing the parallel efficiency are also presented. 
We find that the parallel efficiency  is determined by three main  factors $-$ the   calculation overhead due to relaxation iterations to correct  boundary events in neighboring  processors, the  (extreme)  fluctuations  in the  number of events per cycle in each processor, and the overhead due to interprocessor communications.  Due to the existence of fluctuations and the requirement of global synchronization,   the SR algorithm does not scale, i.e. the parallel efficiency decreases logarithmically as the number of processors increases.
The dependence of the parallel efficiency on simulation parameters such as the processor size, domain decomposition geometry,  and the ratio $D/F$ of  the monomer hopping rate $D$  to the  deposition rate $F$  is also discussed.    
\end{abstract}
\pacs{ 81.15.Aa, 05.10.Ln, 05.10.-a, 89.20.Ff}

\maketitle
\section {Introduction}

Kinetic Monte Carlo (KMC) is an extremely efficient method \cite{bkl, voter, maksym,fichthorn, blue}  
to carry out dynamical simulations when the relevant activated atomic-scale processes are known, and KMC simulations have been    used to model a   variety of dynamical processes ranging from catalysis to thin-film growth.  The basic principle of kinetic Monte Carlo is that the probability that a given event will be  the next event to occur is proportional to the  rate for that event.  Since all processes are assumed to be independent Poisson processes, the time of the next event is determined by  the total overall rate for all processes, and  after each event the rates for all processes are updated as necessary. 

 In contrast to Metropolis Monte Carlo, \cite{Metropolis} in which each  Monte Carlo  step corresponds to a  configuration-independent  time interval and  each event is selected randomly but only accepted  with a configuration-dependent   probability, in kinetic Monte Carlo both the selected event and the time interval between events are configuration-dependent while  all attempts are accepted.   In the context of traditional   Monte Carlo simulations this is sometimes referred to as the n-fold way. \cite{bkl}  While KMC requires additional book-keeping to keep track of the rates of  all possible events,   the  KMC  algorithm is  typically significantly  more efficient than the Metropolis algorithm since no selected moves are rejected.  In particular, for problems such as thin-film growth in which the possible rates or probabilities for events can vary by several orders of magnitude,   the kinetic Monte Carlo algorithm can be orders of magnitude more efficient than Metropolis Monte Carlo.

Because the attempt time in Metropolis Monte Carlo is independent of system configuration,  
parallel Metropolis Monte Carlo simulations   may be carried out by using an asynchronous ``conservative" algorithm.\cite{chandy, Lubachevsky1, Korniss1, Korniss2, Korniss3}  In such an algorithm all processors    whose next attempt time is less than their neighbor's next attempt times are allowed to proceed.  Unfortunately    such a ``conservative" algorithm does not work  for kinetic Monte Carlo since in KMC  the event-time depends on the system configuration.   In particular, since fast events may ``propagate" across processors, the time for an event already executed by a processor may change  due to earlier events in nearby processors, thus leading to an incorrect evolution.  As a result, the development of efficient parallel algorithms for kinetic Monte Carlo simulations remains a  challenging problem.

Lubachevsky has developed\cite{Lubachevsky1} and Korniss et al have implemented\cite{Kornissjcp} a  more efficient version of the  conservative asynchronous algorithm for parallel dynamical Monte Carlo simulations  of the spin-flip Ising model.  The basic idea is to apply   Metropolis dynamics to events on the boundary of a processor, but to accelerate interior moves by using  the $n$-fold way.  The choice of a boundary or interior move is determined by the ratio of the number of boundary sites to the sum of the acceptance probabilities for all interior moves.  While all ``n-fold way" interior moves are immediately accepted, all Metropolis attempts must wait until the neighboring processor's next attempt time is later before being either accepted or rejected. Since such an algorithm is equivalent to the conservative Metropolis Monte Carlo algorithm described above, it  is generally scalable,\cite{Korniss1, Korniss2, Korniss3}  and  has    been found to be relatively efficient in the context of kinetic  Ising model simulations in the metastable regime.   
 \cite{Kornissjcp, Kornisse1,Kornisse2}

Recently we have shown\cite{yshimnovotny}   that such an approach can  be generalized in order to carry out parallel KMC simulations.In this  approach,  all possible KMC moves are first  mapped to Metropolis moves with an acceptance probability for each event given by the rate for that event divided by the fastest possible rate in the KMC simulation.  
At each stage, the choice of a boundary move versus an interior move is determined by the ratio of a fixed number corresponding to the sum of the rates for all {\it possible} events which might occur in the boundary region to the sum of the   rates for all  existing  interior moves.  However, because of the possibility of  significant rejection of boundary events, the parallel efficiency  of such an algorithm can be very low for problems with a wide range of rates for different processes.  For example, we have recently\cite{yshimnovotny}  used  such a mapping to carry out parallel KMC simulations  of a simple 2D solid-on-solid ``fractal" model of submonolayer growth with a moderate   ratio $D/F = 10^5$ of monomer hopping rate $D$ to (per site) deposition rate $F$.   However, due to the existence of significant rejection of boundary events, very low parallel efficiencies were  obtained.\cite{yshimnovotny} Furthermore, in order to use such an approach,  in general one needs to know in advance {\it all}  the possible events and their rates and then to map them to Metropolis dynamics so that all events may be selected with the appropriate probabilities.  While such a mapping may be carried out for  the simplest models, for more complicated models it is likely to be prohibitive.

In order to overcome these problems, we have recently  proposed a semi-rigorous synchronous sublattice (SL) parallel algorithm\cite{yshimsublattice}  in which  each processor's domain is further divided into   sublattices in order to avoid a possible conflict between processors. At the beginning  of a cycle,  one sublattice is randomly selected so that all processors operate on the same sublattice.  Each processor then carries out KMC events for the selected sublattice  over a  time interval which is typically smaller than the inverse of the fastest single-event rate.  At the end of each cycle, each processor communicates with its neighboring processors in order to update its boundary region.  By carrying out extensive simulations of simple models of thin-film growth we have found that this algorithm leads to  a relatively high parallel efficiency and is scalable, i.e. the parallel efficiency is constant as a function of the number of processors $N_p$. 
 However, for extremely small processor sizes (smaller than a typical diffusion length in epitaxial growth) weak finite-size effects are observed. Thus  for problems in  which the diffusion length is large or very small processor sizes are required, it may be preferable to use a more rigorous algorithm.

In this paper we discuss the application of a second rigorous algorithm, the  synchronous relaxation (SR) algorithm,\cite{Eick, Lubachevsky2}    to kinetic Monte Carlo simulations. This algorithm was originally used by  Eick et al \cite{Eick}  to simulate large circuit-switched communication networks.  More recently an estimate of its efficiency has been carried out by   Lubachevsky and Weiss\cite{Lubachevsky2} in the context of Ising model simulations,    In the SR algorithm, all processors remain globally synchronized at the beginning and end of a time interval, while an iterative relaxation method is used to correct errors due to neighboring processors' boundary events.  Since this algorithm is rigorous,  the cycle length can be tuned to optimize the parallel efficiency and several optimization methods are discussed.  However, we find that the requirement of global synchronization leads to a logarithmic increase with increasing processor number  in both the relevant  fluctuations in the number of events per processor as well as  the global communication time. 
 Accordingly, the SR algorithm does not scale since the parallel efficiency  decreases logarithmically as the number of processors increases.  As a result, the parallel efficiency is generally  significantly smaller than for the SL algorithm.

The organization of this paper is as follows.  In Section II we describe the algorithm and discuss  several different methods of optimization.  In Section III we present results obtained using this algorithm for three different models of thin-film growth, along with a brief comparison with   serial results.  We then discuss the three key factors$-$number of additional iterations,   fluctuations, and communications time$-$which determine the parallel efficiency of the SR algorithm. 
 The dependence of the parallel efficiency on  such parameters as the number of processors as well as the cycle length, processor size, and ratio $D/F$ of monomer hopping rate $D$ to (per site) deposition  rate $F$ is also discussed. 
 Finally, in Section IV we summarize our results.

\section {Synchronous Relaxation (SR) Algorithm}

\begin{figure}[]
\includegraphics [width=4.5cm] {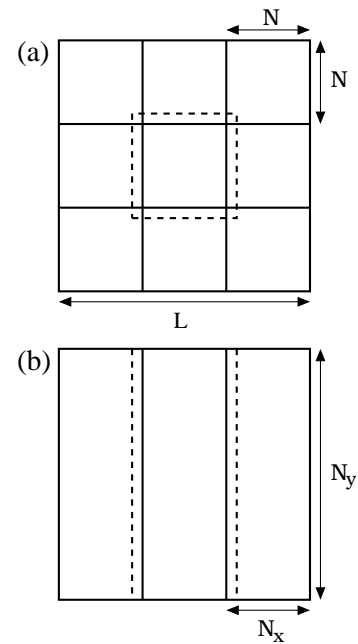}
\caption{\label{fig:Fig1}  {Schematic diagram  of (a) square  and (b) strip decompositions.  Solid lines correspond to processor domains while dashed lines  indicate ``ghost-region" surrounding central processor. }}
\end{figure}

As in previous work on  the ``conservative" asynchronous algorithm,\cite{Lubachevsky1, Korniss1} in  the synchronous relaxation (SR) algorithm, different parts of the system are assigned  to different processors  via spatial decomposition. For the thin-film growth simulations considered here, one may consider two possible  methods of spatial decomposition, a square decomposition and  a strip decomposition as shown in Fig. \ref{fig:Fig1}. Since the square decomposition requires communications with 4 neighbors while the strip decomposition only requires  communications with 2 neighbors,   we expect that  the strip decomposition will have  reduced communication overhead. Accordingly,  all the results presented here are for the case of  strip   decomposition.   However, as discussed in more detail later, there may be some cases where the square decomposition is preferable. 

In order to avoid communicating with processors beyond the nearest-neighbors, the processor size must be  larger than the range of interaction (typically only a few lattice units in simulations of thin-film growth).  
In addition, in order for each processor to calculate its event rates, the configuration in neighboring processors must be known as far as the range of interaction.  As a result, in addition to containing the configuration information for its own domain, each processor's array also contains a ``ghost-region" which includes the relevant information about the neighboring processor's configuration beyond the processor's  boundary.

\begin{figure}[]
\includegraphics  [height=4.0cm] {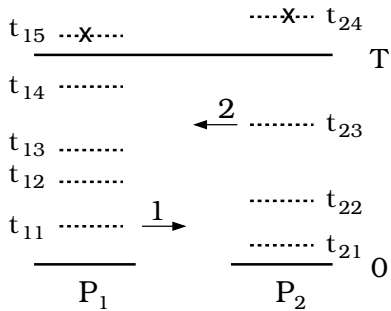}
\caption{\label{fig:Fig2}  {Diagram showing time evolution in the synchronous relaxation algorithm.  Dashed lines  correspond to   events, while dashed line with an $\mathsf{X}$ corresponds to an event which is rejected since it exceeds the time interval $T$. Arrows indicate boundary events carried out by processors $P_1$ and $P_2$.}}
\end{figure}

We now describe the SR algorithm in detail. At the beginning of each   cycle corresponding to a time interval $T$,  each processor initializes its time to zero.   A first iteration is then performed  in which each processor carries out KMC events  until the time of the next event exceeds  the time interval $T$ as shown in Fig. \ref{fig:Fig2}. 
As in the usual serial KMC, each event is carried out with  time increment $\Delta t_i = - \ln(r_i)/R_i$  where $r_i$ is a uniform random number between $0$ and $1$ and $R_i$ is the total KMC event rate.  
At the end of each iteration, each processor   communicates any boundary events with its neighboring processors, i.e. any events which are in the range of interaction of a neighboring processor and which could thus potentially affect the neighboring processor's event rates and times. Here we define an event   as consisting of the lattice sites which have changed due to the event along with the unique time $t_i$ of the event.  
If at the end of a given iteration, a processor has received any {new} or missing boundary events (i.e. any  boundary events different from those received in the previous iteration) then that processor must ``undo" all of the KMC events  which occurred after the time of the earliest  new  or missing boundary event, and then perform another iteration starting at that point  using the new boundary information received.  
However, if   no processors have received new or missing boundary events, then the iterative relaxation is complete, and all processors move on to the next cycle. 
In order to check for this, a global communications between all processors is performed at the end of each iteration.

In order for the iteration process to converge, one must ensure that within the same cycle the same starting configuration always leads to the same event or transition.  Since pseudorandom numbers are used in  the KMC simulations considered here, this requires keeping a list of all the random numbers used during that cycle, and backtracking appropriately along the random number  list as events are ``undone" so that the same random numbers are used for the same configurations.   
In the ideal implementation described above,  events are only redone starting from the earliest new boundary event.  However,  in the KMC simulations carried out here, lists were used to efficiently select and keep track of all possible events.  Since properly undoing such lists is  somewhat complex,  here we have used a slightly less efficient but  simpler method in which every iteration was restarted at the beginning of the cycle. In this case, the  necessary changes in the configuration and random numbers were ``undone" back to the beginning of the cycle, while the state of the lists at the beginning of the  cycle was restored.  Since there is significant overhead associated with ``undoing" each move, and since  in every iteration except the first, one needs to ``undo" on average only half of the events in the previous iteration  we estimate such a simplification leads to at most a 25\% reduction in the parallel efficiency.

We now consider the general dependence of the parallel efficiency   on the cycle time $T$.  If the cycle time is too short  then there will be a small number of events in each cycle and as a result there will be  large fluctuations in the number of events in different processors.  This  leads to poor utilization, i.e. some processors may process events during a given cycle while others may have very few or no events.  In addition, for a short cycle time the communication latency may become  comparable to  the calculation time  which also leads to a  reduction in the parallel efficiency.  On the other hand, a very long cycle time  will lead to a large number of  boundary events in each cycle, and as a result the number of relaxation iterations will be large.  Thus, in general the cycle length $T$ must be optimized in order to balance out the competing effects of communication latency, fluctuations,  and iterations in order to obtain the maximum possible efficiency.

 We have used three different methods to control the time interval $T$ in order to optimize the parallel efficiency. In the first method, we have used  a fixed cycle length  (e.g. $T=\phi/D$ where $D$ is the monomer hopping rate)   and then carried out simulations with different values of $\phi$  in order to determine the optimal cycle length and   maximize the parallel efficiency.  In the second method, we have used feedback to dynamically control the cycle length $T$ during a simulation. In particular,   every $3 - 10$ cycles corresponding to a feedback interval,  the  elapsed execution time was either calculated or measured, and then used to calculate the ratio $\rho$ (proportional to the parallel efficiency) of the average number of events per cycle  $n_{av}$ to the execution time.  Based on the values of $\rho$ obtained during the previous two feedback intervals, the cycle length $T$ was  adjusted in order to maximize the parallel efficiency.    In the third  optimization method  the cycle length was dynamically controlled in order to attain a pre-determined value for a target quantity  such as the number of events per cycle ($n_{opt}$)  or the number of iterations per cycle ($n_{it}$) whose optimal value was  determined in advance. This method turned out be the most effective since the parallel performance    depends strongly on the number of  iterations and/or the number of events per cycle. In contrast,   while the parallel efficiency obtained using direct feedback was significantly better than that obtained using the first method, it was not quite as good as that obtained using the third method described above.
 As a result, here we focus  mainly on the first and last methods. 
Unfortunately,  both of these methods require the use of additional simulations in order to determine the 
optimal parameters.

\section {Results}

\begin{figure}[]
\includegraphics  [height=3.5cm] {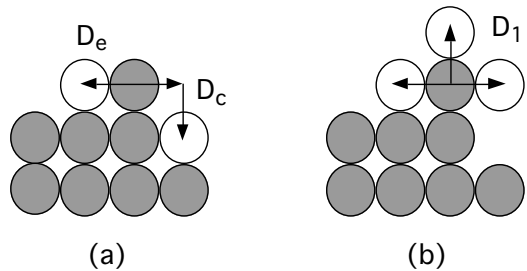}
\caption{\label{fig:Fig3}  {Schematic diagram of  island-relaxation mechanisms for (a) edge-and-corner  and (b)  reversible models.}}
\end{figure}

In order to test the performance and accuracy of the synchronous relaxation  algorithm we have used it to simulate three specific models of thin-film growth. In particular, we have studied  three  solid-on-solid (SOS)   growth models on a square lattice:  a ``fractal" growth model, an edge-and-corner diffusion (EC) model, and a reversible model with one-bond detachment (``reversible model").  In each of these three models the lattice configuration is represented by a two-dimensional array of heights and periodic boundary conditions are assumed.  In the ``fractal" model,\cite{islandprb}  atoms (monomers) are deposited onto a square lattice with (per site)  deposition rate $F$, diffuse (hop) to nearest-neighbor sites with hopping rate $D$ and attach irreversibly  to other monomers or clusters via a nearest-neighbor bond (critical island size of $1$). The key parameter  is the ratio $D/F$  which is typically much larger than one in epitaxial growth.   In this model fractal islands are formed in the submonolayer regime  due to the absence of island relaxation. The EC model is the same as the fractal model except that island relaxation is allowed, i.e.   atoms which have formed a single nearest-neighbor bond with  an  island may diffuse along the  edge of the island with diffusion rate $D_e = r_e D$ and around island-corners with rate $D_c = r_c D$ (see Fig. \ref{fig:Fig3}).  Finally, the reversible model is  also similar to the fractal model except that  atoms with one-bond (edge-atoms) may hop along the step-edge or away from the step with rate $D_1 = r_1 D$, thus allowing both edge-diffusion and single-bond detachment.  For  atoms hopping up or down a step, an extra Ehrlich-Schwoebel barrier   to interlayer diffusion \cite{ES}  may also be included.  In this model, the critical island size  $i$\cite{Venables} can vary from $i = 1$ for small values of $r_1$,  to $i = 3$ for  sufficiently large values of  $D/F$ and  $r_1$.\cite{islandprl}

 For the fractal and reversible models, the range of interaction corresponds to one nearest-neighbor (lattice) spacing, while for the EC model it corresponds to the next-nearest-neighbor distance.  Thus, for these models  the width of the ``ghost-region"  corresponds to one   lattice-spacing.  We note that at each step of the simulation, either a particle is   deposited  or a particle  diffuses  to a nearest-neighbor or next-nearest-neighbor lattice site.  
 In more general models, for which concerted moves involving several atoms may occur, \cite{Zhangdimer, Hamilton, Jonsson, tad4} the ghost region needs to be at least as large as the range of interaction and/or the largest possible concerted move.  In such a case, the processor  to which a concerted event belongs can be determined by considering the location of the center-of-mass of the atoms involved in the concerted move. 
 
 In order to maximize both the serial and parallel efficiency  in our KMC simulations,  we have used  lists to keep track of all possible events of each type and rate.  Each processor maintains  a  set of lists which contains  all possible moves of each type.  A binary tree  is used to select which type of move will be carried out, while the particular move is then randomly chosen from the list of the selected type.  After each move, the lists are updated. 

\subsection {Computational Details}

In order to test our algorithm we have carried out both ``serial emulations" as well as    parallel simulations. However, since our main goal is to test the  performance and scaling behavior on parallel machines we have primarily focused on direct parallel simulations using the Itanium and AMD clusters at the Ohio Supercomputer Center (OSC) as well as on the Alpha cluster at the Pittsburgh Supercomputer Center (PSC).  All of these clusters have fast communications$-$the Itanium and AMD clusters have Myrinet and the Alphaserver cluster has Quadrics. In our simulations, the interprocessor communications  were carried out using MPI (Message-Passing Interface).

\subsection {Comparison with Serial Results}
 
\begin{figure}[]
\includegraphics [width=7.0cm] {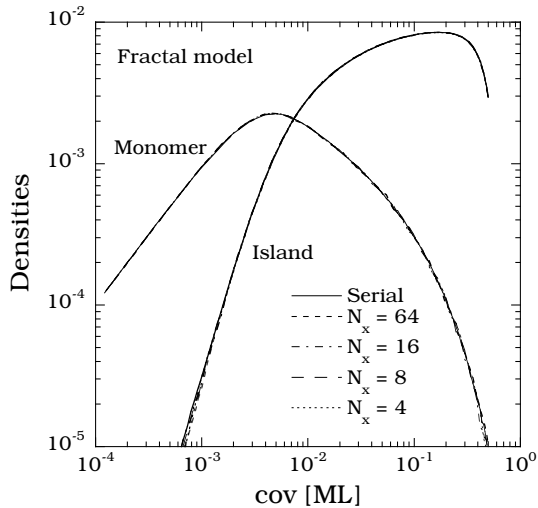}
\caption{\label{fig:Fig4}  {Comparison between serial and parallel results  for the fractal model with $L=256$ and $D/F = 10^5$.}}
\end{figure}

We first present a brief comparison between serial  and parallel results 
 in order to verify the correctness of our implementation of the SR algorithm. Figure \ref{fig:Fig4} shows the monomer and island densities as a function of coverage $\theta\leq 0.5$ for the fractal model with $D/F=10^5$ and system size $L=256$ with parallel processor sizes $N_y=256$ and $N_x=4, 8, 16$ and $64$ corresponding to $N_p=64, 32, 16$ and $4$ respectively (where $N_p$ is the number of processors).  As can be seen, there is excellent agreement between the serial and parallel calculations even for $N_x=4$, the smallest processor size we have tested. Using the SR algorithm, we have also obtained excellent agreement  between serial and parallel results for the monomer and island densities for the EC model (not shown).

\subsection{Calculation of parallel efficiency}

The parallel efficiency is determined by the competing effects of communication time, fluctuations, and number of relaxation iterations. In particular, the parallel execution time $t_{N_p} (\tau)$ for $N_p$ processors in  cycle $\tau$ can be written as
\begin{equation}
t_{N_p} (\tau) = t_{calc} (\tau) + t_{com} +  t_{other}, \label {eqn1}
\end{equation} 
where $t_{calc}(\tau)$ and  $t_{com}$ denote the calculation  and communication time respectively,  while the last term $t_{other}$  includes all other  timing costs not included in $t_{calc}$   such as  sorting boundary events received from  neighbors  and comparing new boundary events with old ones to see if a new iteration is needed. If there are few  boundary events (as is often the case), then $t_{other}$ may be ignored. 
The calculation time $t_{calc}(\tau) $ in Eq.~\ref{eqn1} may be written as, 
\begin{equation}
t_{calc} (\tau)  =  t^{1}_{KMC}\times (n_{av}(\tau) +  \Delta(\tau) )  \label {eqn2} 
\end{equation}
where  $t^{1}_{KMC}$ denotes the average serial calculation time per KMC event, $n_{av}(\tau)$ is the average number of actual events (averaged over all processors) per processor in  cycle $\tau$, and $\Delta(\tau) $ corresponds to the additional number of events which must be processed due to fluctuations and relaxation iterations. 

 Since all processors are synchronized after each iteration, in each iteration the total calculation time is determined by the processor which has the maximum number of events to process.  Therefore  one may write,
\begin{equation}
\Delta(\tau) =  (\sum_{j=1}^{I}[ n'_{\max}(\tau,j)+\lambda\ n'_{\max}(\tau,j-1)]) - n_{av}(\tau) \label{eqn3}
\end{equation}
where  $n'_{\max}(\tau,j)$ is the maximum (over all processors) number of new events in the $j$th iteration, $I$ is the total number of  relaxation  iterations, and $\lambda \simeq 0.14$ is a factor which reflects the reduced work to ``undo" a KMC event as compared to executing a KMC event.  Thus, the parallel efficiency (PE) can be approximated as
\begin{equation}
PE = \frac{t_{1p}}{\langle t_{N_p}(\tau) \rangle}  \simeq  \left[ 1 + \frac{\langle\Delta(\tau)\rangle}{n_{av}} + \frac{{t}_{com}}{t_{1p}}\right]^{-1}  \label{eqn4}
\end{equation}
where $t_{1p}=n_{av}\times t^{1}_{KMC}$  is the time for a   serial simulation of a single processor's domain,  $n_{av} = \langle n_{av} (\tau) \rangle$, and the brackets denote an average over all cycles.   If the communication time is negligible compared to $t_{1p}$ (i.e., $t_{com}/t_{1p}\rightarrow 0$), the maximum possible parallel efficiency can be approximated as 
\begin{equation}
PE =   \left[ 1 + \frac{\langle\Delta(\tau)\rangle}{n_{av}}\right]^{-1}\label{eqn5}
\end{equation}
We note that $\langle \Delta (\tau) \rangle$  depends primarily on two quantities, the (extreme)  fluctuations over all processors in the number of actual events  in each cycle $\Delta n^e/n_{av} \equiv (n_{\max} - n_{av})/n_{av}$, and the average number of iterations $I$ per cycle. In particular, one may approximate the additional overhead due to iterations and fluctuations as,
\begin{equation}
 \frac{\langle\Delta(\tau)\rangle}{n_{av}} \simeq    \eta (I - 1) (1 +  \frac  {\Delta n^e} {n_{av}} ) \label{eqn6}
 \end{equation} 
where a factor of  $\eta$ with $\eta \le 1$ has been included to take into account the fact that after the first iteration, the number of new events $n'_{\max}$ is  typically  less than $n_{\max}$.  This result indicates that  in the limit of negligible communications overhead, both the average number of iterations per cycle  $I$ and the relative fluctuations $\Delta n^e/n_{av}$ should be small in order to maximize the parallel efficiency. 
 We now consider the dependence of each of these quantities on the the cycle length $T$ and the number of processors $N_p$.

\subsection{Number of iterations}

\begin{figure}[]
\includegraphics [width=7.0cm] {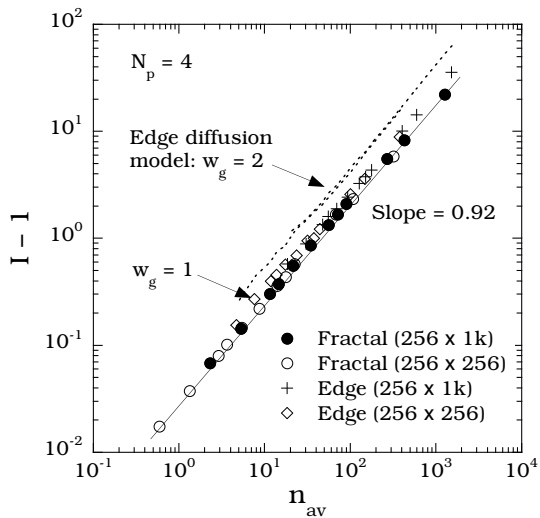}
\caption{\label{fig:Fig5}  {Number of additional iterations as a function of average number of events per cycle $n_{av}$ with $T=1/D$. For the EC model $r_e=0.1$ and $r_c=0$ are used with $w_g=1$ and $2$. Here $N_p=4$ and $\theta=1\ ML$ for all cases and $D/F=10^3-10^7$.   }}
\end{figure}

Figure \ref{fig:Fig5} shows the  number of additional iterations  beyond the first iteration $I' = I-1$ as a function of the  average  number of events per cycle $n_{av}$  for the fractal and EC models with  $N_p = 4$ using  strip decomposition and  two different processor sizes for different values of $D/F$.  
Also shown in Fig. \ref{fig:Fig5} are results for a larger than required ghost-region $w_g = 2$ in order to test the dependence of the number of iterations on the range of interaction. As can be seen,  the number of additional iterations  is roughly  linearly proportional to the average number of events per cycle.  Interestingly,   for the  same  average number of events per cycle $n_{av}$,  the number of additional iterations depends relatively weakly on the model, the processor height $N_y$, and the value of $D/F$.   However,  doubling  the width of the ``ghost" region from $w_g  = 1$ to $w_g = 2$ leads to an increase by a factor of approximately $1.5$ in  the number of iterations.

\begin{figure}[]
\includegraphics [width=7.0cm] {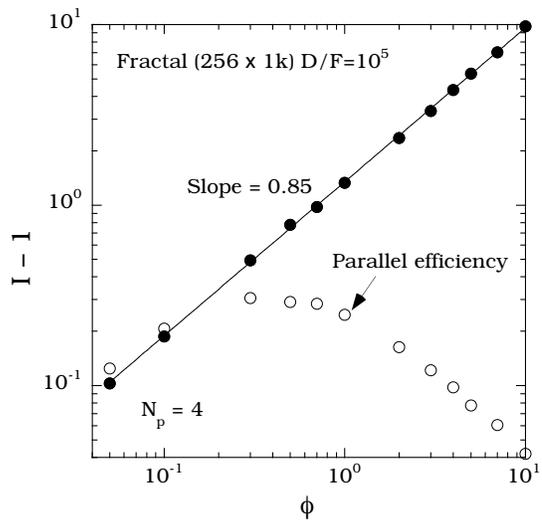}
\caption{\label{fig:Fig6} {Number of additional iterations and parallel efficiency for the fractal model as a function of the multiplication factor $\phi$ with $N_p=4$, $\theta=1\ ML$ and $T=\phi/D$.}}
\end{figure}

Figure \ref{fig:Fig6}   shows  the number of additional iterations $I'$ as a function of the cycle length    $T$ for  the fractal model with $D/F=10^5$, $N_p=4$, and $N_x = 256, N_y = 1024$.  As can be seen, the number of additional iterations  is roughly  but not quite proportional to the cycle length $T$.  Also shown in Fig.~ \ref{fig:Fig6} is the parallel efficiency, which has been directly measured from the execution time using the definition given in Eq.~\ref{eqn4}. 
 The maximum parallel efficiency occurs when $T = T_{opt} \simeq 3.0\times 10^{-6}$ and corresponds to $n_{av} \simeq 30$ KMC events per cycle.  We   have also calculated (not shown)  the optimal cycle length $T_{opt}$ for the same processor size for  other values of $D/F$ ranging from  $10^3$ to  $10^7$.  While  the   optimal cycle length varies by approximately two orders of magnitude over this range  of $D/F$,  the   average optimal number of  events per cycle does not change much$-$$n_{opt}\simeq 38$ and $30$ for $D/F=10^3$ and $10^7$  respectively.

\begin{figure}[]
\includegraphics [width=7.0cm] {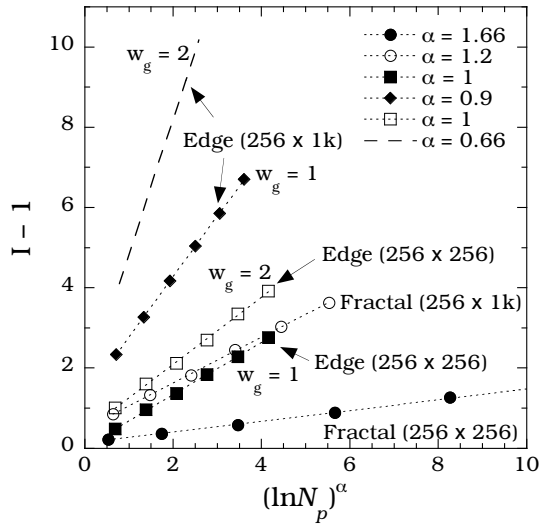}
\caption{\label{fig:Fig7}  {Number of additional iterations as a function of the  number of processors $N_p$ with $2 \le N_p \le 64$  for the fractal and edge diffusion models with $D/F=10^5$, $T=1/D$ and $\theta=1\ ML$. In the EC model, $r_e=0.1$ and $r_c =0$ and the width of the ghost region $w_g=1$ unless specified.  }}
\end{figure}

Since the probability of  an ``extreme" number of boundary events in one of the  processors increases with   the number of processors for fixed processor size, the number of iterations increases with $N_p$. As shown in Fig.~\ref{fig:Fig7}, such an increase is well described by the logarithmic  form, $I-1= a_0(\ln{N_p})^{\alpha}$
where the exponent  $\alpha$ ranges from $0.66$ to $1.7$ depending on the model and  processor size.  
Fig.~\ref{fig:Fig7} also indicates that an increase of the interaction length from $w_g=1$ to $w_g=2$ also yields an approximate doubling in the number of iterations when a fixed time interval $T=1/D$ is used.  
Thus,   in order to keep the number of iterations constant, the time interval must decrease as the range of interaction increases.

\subsection{Fluctuations in number of events}

\begin{figure}[]
\includegraphics [width=7.0cm] {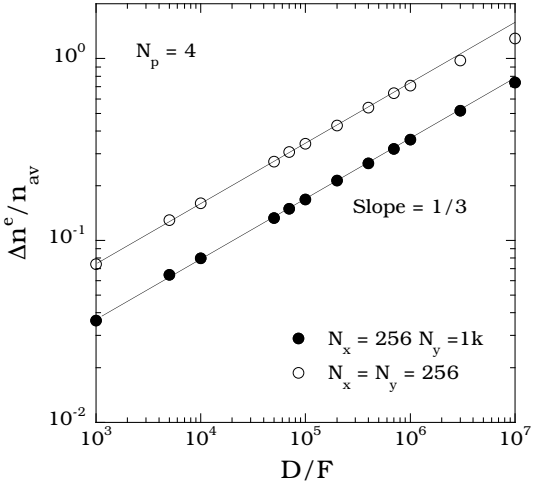}
\caption{\label{fig:Fig8}   {Relative fluctuation in number of events for the fractal model as a function of $D/F$ with $T=1/D$,  $N_p=4$ and $\theta=1\ ML$.}}
\end{figure}

A second important factor which determines the parallel efficiency is  the  existence of fluctuations in the number of events in different processors. In particular, since all processors are globally synchronized,  the processor having the maximum number of events  $n_{\max}$  can  determine the execution time of each iteration.  
Thus the extreme   fluctuations $\Delta n^e/n_{av}$, as opposed to the usual r.m.s.  fluctuations, determine the parallel efficiency.  

\begin{figure}[]
\includegraphics [width=7.0cm] {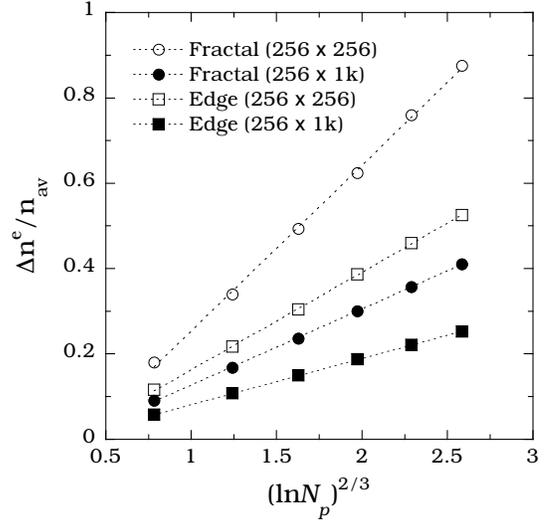}
\caption{\label{fig:Fig9}  {Relative fluctuation in number of events for fractal and edge diffusion models as a function of  number of processors $N_p$ with $D/F=10^5$, $\theta=1\ ML$ and $T=1/D$, where $w_g=1$ in all cases. In the EC model, $r_e=0.1$ and $r_c=0$. Dotted lines are linear fits to data.}}
\end{figure}

Figure \ref{fig:Fig8} shows the measured fluctuations  $\Delta n^e/n_{av}$  for the simple fractal model as a function of  $D/F$ for fixed processor size $N_x = 256$, $N_y = 1024$ and $N_p = 4$ averaged over many cycles.   As expected the relative fluctuations in a smaller system are larger than those in a bigger system.  For the simple fractal model,   one expects  that $n_{av} \sim N_1  \sim (D/F)^{-2/3}$ which implies $\Delta n^e/n_{av} \sim 1/\sqrt{n_{av}}\sim (D/F)^{1/3}$. 
 As can be seen,  there is very good agreement with this form for the $D/F$-dependence.

Figure \ref{fig:Fig9} shows the relative (extreme)  fluctuations as a  function of the number of processors $N_p$ for the fractal and EC models with two different processor sizes. As for the dependence of the number of iterations on $N_p$, we find a logarithmic dependence.  In particular, we find that $\Delta n^e/n_{av} \sim (\ln{N_p})^{\gamma}$ with $\gamma=2/3$ regardless of model and processor size. Again, for a fixed $N_p$, a bigger system shows smaller relative  fluctuations than a smaller system.  In addition, for the same processor size, the EC model shows smaller relative  fluctuations than the fractal model due to the additional number of edge-diffusion events   in the model.

\begin{figure}[]
\includegraphics [width=7.0cm] {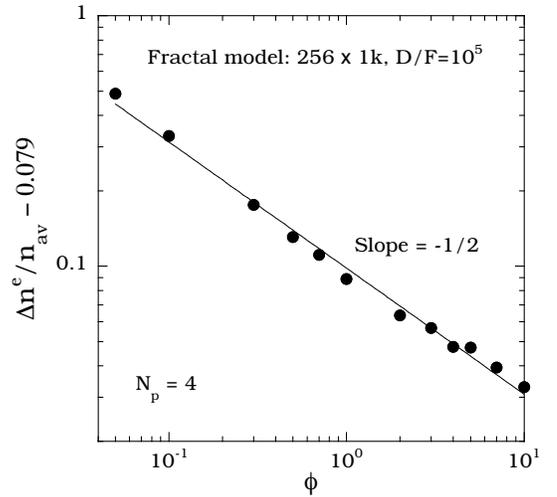}
\caption{\label{fig:Fig10}  {Power-law decay in the relative fluctuations for the fractal model as a function of multiplication factor $\phi$ and $T=\phi/D$ with $N_p=4$ and $\theta=1\ ML$. }}
\end{figure}

We now consider the dependence of the  fluctuations   on the time interval $T$.   For the fractal model, the average number of events per cycle in each processor may be written, $n_{av} = N_x N_y (F + N_1 D) T$ where $F$ is the deposition rate,  $D$ is the monomer hopping rate, and $N_1$ is the monomer density.  The fluctuation in the number of events may be written as the sum of the  fluctuation (proportional to $n_{av}^{1/2}$)  assuming all processors have the same average event rate,  and  an additional  term due to fluctuations in the number of monomers in different processors, i.e. $\Delta n^e  \sim n_{av}^{1/2} + N_x N_y ~\delta N_1^e ~D  T$. We note that the fluctuation $\delta  N_1^e = N_1^{\max} - \langle N_1 \rangle $ also depends on  the number of processors $N_p$ and the processor size $N_x N_y$.  Dividing to obtain the (extreme) relative fluctuation we obtain,
\begin {equation}
\frac {\Delta n^e} {n_{av}} = \frac {(D/F)~ \delta N_1^e} {1 + (D/F) \langle N_1 \rangle}  + [ N_x N_y (F/D  + \langle N_1 \rangle) \phi]^{-1/2}
\end{equation}
where $\phi =  DT$.  The first term is independent of the cycle length $T = \phi/D$   while for $D/F >> 1$,  $\langle N_1 \rangle  >> F/D$ and so the second term is simply proportional to $\phi^{-1/2}$. 
As can be seen in Fig. \ref{fig:Fig10}, we find good agreement with this form for the fractal model with $N_p = 4$, $D/F = 10^5$, $N_x = 256, N_y = 1024$ and the time interval $T$ ranging over more than  two decades. 

\subsection{Communication time and event optimization}

The third factor which determines the parallel efficiency is the communications overhead. For the case of strip decomposition, in every iteration each processor must carry out  two local send/receive communications with its neighbors.  Typically, a send/receive communication with a small  message size ($<100$ bytes) between two processors in the same node takes less than 10 $\mu s$ but it can take  longer  if they are in different nodes. For a larger message size the communication overhead increases linearly with message size. Since the processor size in all of our simulations is moderate, the message size is only about 2000 bytes which takes roughly $30 \mu s$. 

A global communication (global sum or ``AND" and broadcast) must also be carried out at the end of every  iteration to check if a new iteration is necessary.  The time for the global sum and broadcast  is     larger than for  a  local   send/receive and increases logarithmically with the number of processors $N_p$. Overall, we find that the estimated minimum total communication overhead per cycle is roughly $60 \mu s$ for a small number of processors.  In comparison, the serial calculation time $t^1_{KMC}$ for one KMC event is  about $5\mu s$ on the Itanium cluster. Thus, even for a small number of processors  the overhead due to communications  ($t_{com}/t_{1p}$) is significant  unless  $n_{av} >> 12$.

\begin{figure}[]
\includegraphics [width=7.0cm] {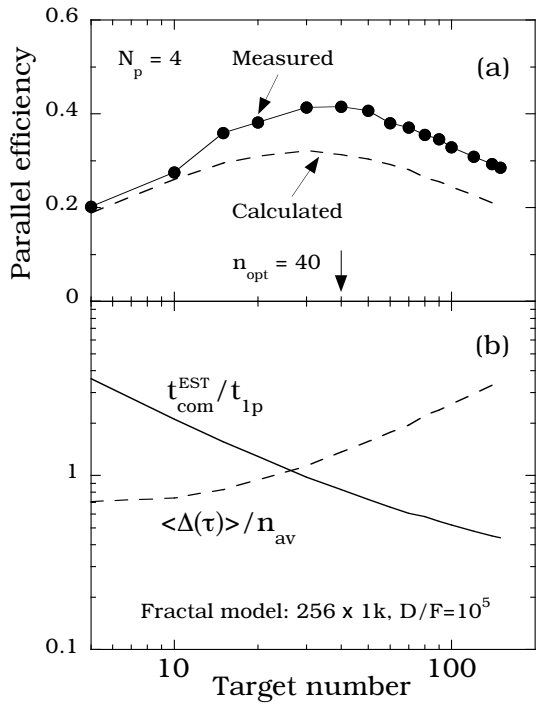}
\caption{\label{fig:Fig11}   {(a) Parallel efficiency and (b) measured additional number of events and estimated communication time per the average number of events as a function of target number of events with $\theta=1ML$ and $N_p=4$. }}
\end{figure}

One way to    maximize the parallel efficiency is to use the event optimization method.  In this method, the cycle length $T$ is  dynamically adjusted  during the course of the simulation  in order to achieve a fixed target number of events per cycle.  By varying the target number of events and measuring the simulation time one may determine the optimal target number $n_{opt}$.  Figure \ref{fig:Fig11} (a) shows the measured  parallel efficiency for the fractal model with $N_p = 4, N_x = 256, N_y =1024$,  and $D/F = 10^5$ as a function of the target number of events.  As can be seen, for a target number given by $n_{opt} = 40$ there is an optimal efficiency of approximately 41\%.  Also shown (dashed line) is the parallel efficiency  calculated using  the measured additional number of events  $\langle \Delta (\tau) \rangle$ due to fluctuations and relaxation iterations along with the estimated communication time  which may be approximated by the fit  $t_{com}^{est}/ t_{1p} \simeq 16 I/n_{av}$.  The resulting calculated parallel efficiency curve is close to the measured curve  but has a  slightly lower peak.  Fig. \ref{fig:Fig11}(b) shows separately  the two contributions to the calculated parallel efficiency  $\langle \Delta (\tau) \rangle/n_{av}$ and  $t_{com}/ t_{1p}$ as a function of the target number of events.  As can be seen, the peak of the parallel  efficiency is close to  the point where these two contributions have   the same magnitude.

\subsection{Parallel efficiency as a function of $D/F$}

\begin{figure}[]
\includegraphics [width=7.0cm] {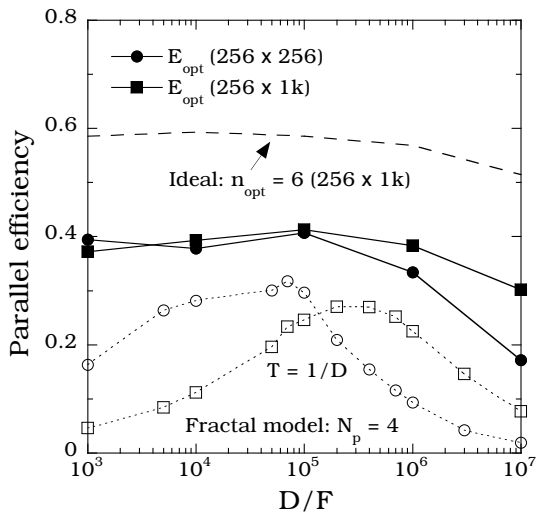}
\caption{\label{fig:Fig12}   {Parallel efficiency for  fractal model as function of $D/F$ with $T=1/D$ (open symbols) and  event  optimization ($E_{opt}$) method (filled symbols). Same symbol shape is used for the same processor size. Here $N_p=4$ and $\theta=1\ ML$ in all cases. In the $E_{opt}$ method, $n_{opt}=18$  for $N_x=N_y=256$ and all $D/F$, and $n_{opt}=50 (40)$ for $N_x=256, N_y=1k$ with $D/F<10^5$ ($\geq 10^5$).}}
\end{figure}

We now consider the parallel efficiency of the SR algorithm as a function of  $D/F$ for the fractal model for a fixed number of processors $N_p = 4$. As shown in Fig.~\ref{fig:Fig12}, when a fixed time interval $T=1/D$ is used, the parallel efficiency (open symbols)  shows  a distinct peak as a function of $D/F$, with a maximum parallel efficiency  $PE \simeq 0.3$ for both processor sizes.  The existence of such a peak may be explained as follows.  For small $D/F$ the PE is low due to the large number of events in each processor which leads to a large number of boundary events and relaxation iterations in each cycle. For large $D/F$ the number of events per cycle is reduced but  the communications overhead and fluctuations become significant due to the small number of events.  At  an intermediate value of $D/F$ which increases with increasing processor size,  neither of these effects dominate  and  the parallel efficiency is maximum.   

In contrast, when the event optimization method is used (filled symbols) the   parallel efficiency is  significantly higher and is almost independent of $D/F$  for $D/F \leq 10^6$.  Although the value of the optimum target number of events  $n_{opt}$ increases with processor size there is only a weak dependence on $D/F$ for fixed processor size.  Also shown in Fig.  \ref{fig:Fig12}  (dashed line) is the estimated ideal parallel efficiency assuming negligible communications overhead.   In this case, a small  target number of events, $n_{opt}=6$ was found to yield the maximum ideal parallel efficiency over the range of $D/F$ studied here.

\begin{figure}[]
\includegraphics [width=7.0cm] {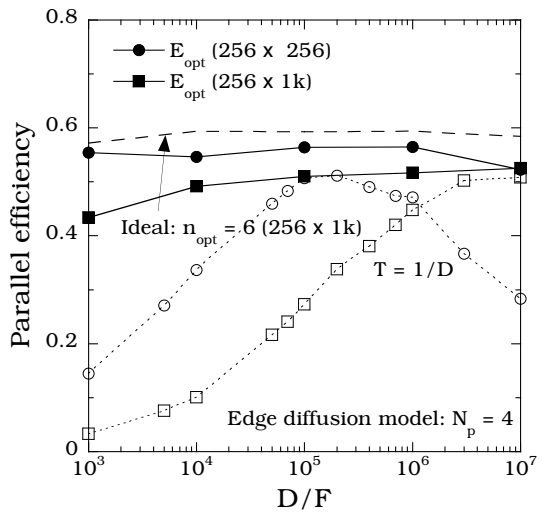}
\caption{\label{fig:Fig13}  {Parallel efficiency for edge diffusion model as  function of $D/F$ with $T=1/D$ (open symbols) and with  event  optimization method (filled symbols). Same symbol shape is used for the same processor size. Here, $N_p=4$, $\theta=1\ ML$, $r_e=0.1$ and $r_c=0$.  In the $E_{opt}$ method, $n_{opt}=23 (30)$ for $N_x=N_y=256$ ($N_x=256$ and $N_y=1k$) for all $D/F$. Dashed line represents  ideal parallel efficiency $1/\left[1+\langle\Delta  (\tau) \rangle/n_{av}\right]$ using $n_{opt}=6$ with $N_x=256$ and $N_y=1k$.}}
\end{figure}

Similar results  are shown in Fig. \ref{fig:Fig13} for the edge diffusion model.  
Since for the edge-diffusion model the ``event density" is significantly higher  than for the fractal model, the communications overhead and fluctuations are  significantly reduced.  
As a result, for the case of  a fixed time interval  $T=1/D$, the maximum parallel efficiency is about $50\%$  for the edge-diffusion model which is significantly higher than  the peak value of $30\%$ for the fractal case. When the event optimization method is used, the PE is also higher than for the fractal case and  is again roughly independent of $D/F$.  Also shown in Fig. \ref{fig:Fig13} for the larger processor size is the calculated ideal parallel efficiency assuming negligible communications overhead and a target number of events given by  $n_{opt} = 6$ (dashed line).   Due to the reduced   communications  overhead for this model, the ideal PE   is only slightly higher than the corresponding optimal  PE with communications included (filled squares).

\subsection{Parallel efficiency as a function of $N_p$}

\begin{figure}[]
\includegraphics [width=7.0cm] {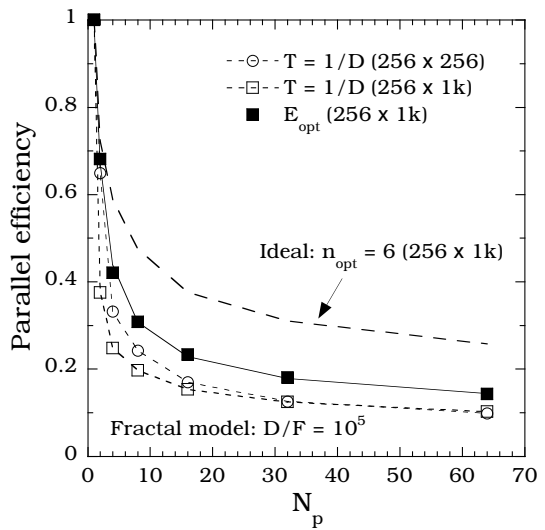}
\caption{\label{fig:Fig14}   {Parallel efficiency for  fractal model  with $D/F =10^5$ as a function of number of processors with $T=1/D$ (open symbols) and with  event optimization method ($n_{opt}=40$). Dashed line represents  ideal parallel efficiency using a target number of events  $n_{opt}=6$. Solid line is a fit to the event-optimization data  with a form, $PE = 1/[1+0.81(\ln{N_p})^{1.4}]$.}}
\end{figure}

We first consider the dependence of the parallel efficiency on the number of processors $N_p$ with fixed processor size. As before, the parallel efficiency  is defined as  the ratio of the execution time for an ordinary serial simulation of one processor's domain to  the parallel execution time of $N_p$ domains using $N_p$ processors (Eq.~\ref{eqn4}).   
Fig. \ref{fig:Fig14} shows the parallel efficiency for the fractal model  with $D/F = 10^5$ as a function of the number of processors $N_p$ for fixed processor size. Results (open symbols) are shown for two different processor sizes for the case of fixed cycle  length $T = 1/D$.  Also shown (filled symbols) is the parallel efficiency obtained using event optimization   for the larger processor size.  While the parallel efficiencies  obtained using event optimization are significantly higher than the corresponding results obtained using  a fixed time interval,  the percentage difference  decreases slightly as the number of processors increases.

The solid line in  Fig. \ref{fig:Fig14} shows a fit of the form 
\begin{equation}
PE = 1/[1+ c~ (\ln N_p)^{\beta}]  \label{fit} 
\end{equation}
 (see Eq.~\ref{eqn4})  to the parallel efficiency obtained for the larger processor size using event optimization.  
As can be seen, there is excellent agreement with the simulation results. The  value of the exponent  ($\beta = 1.4$) is in reasonable  agreement with the dependence of the number of additional iterations on $N_p$ shown in Fig.~\ref{fig:Fig7}.   Also shown in Fig. \ref{fig:Fig14} is the ideal parallel efficiency in the absence of communication overhead calculated using a  target number of events given by   $n_{opt} = 6$.  
As expected, the ideal PE  is significantly larger than the actual PE even for large $N_p$.  In this case a similar fit of the form of Eq.~\ref{fit} may be made but with $\beta \simeq 1.1$.

\begin{figure}[]
\includegraphics [width=7.0cm] {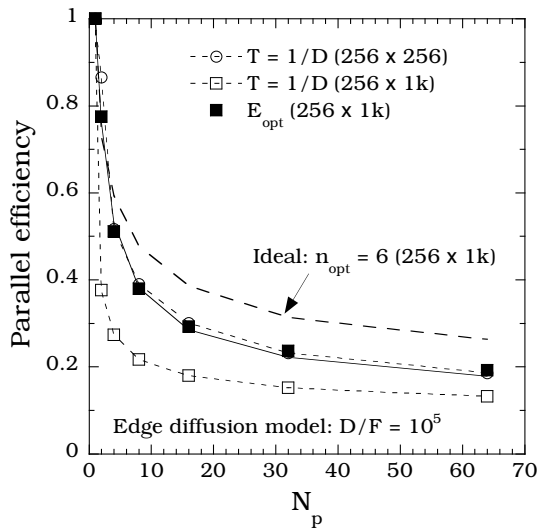}
\caption{\label{fig:Fig15}  {Parallel efficiency for  edge diffusion model with $D/F = 10^5$ as a function of $N_p$ with $T=1/D$ (open symbols) and with  event optimization method ($n_{opt}=30$). Dashed line represents  ideal parallel efficiency obtained using a  target number of events given by $n_{opt}=6$. 
Solid line is a fit to the data for $E_{opt}$ with a form, $PE = 1/[1+0.54(\ln{N_p})^{1.5}]$. }}
\end{figure}

Fig. \ref{fig:Fig15} shows similar results for the edge-diffusion model with $D/F = 10^5$, $D_e = 0.1 D$ and $D_c = 0$.  For the larger processor size ($N_x = 256, N_y = 1024$) both  the results obtained using  a fixed cycle size and those  using event optimization   are very similar to the corresponding results already obtained for the fractal model.  However, for a  fixed cycle length the parallel efficiencies for the  smaller processor size ($N_x = N_y = 256$) are somewhat higher than the corresponding results for  the fractal model.  Again, the ideal parallel efficiency is well described by a fit of the form of Eq.~\ref{fit} with $\beta \simeq 1.1$.   In general, all the parallel efficiencies shown in Figs. \ref{fig:Fig14} and \ref{fig:Fig15} are reasonably well described by  fits of the form of   Eq.~\ref{fit}   with  $0.66\leq \beta \le 1.5$.

We now consider  the dependence of the  parallel efficiency on the number of processors for a fixed  total system size $L$ in the case of strip geometry  (i.e., $N_y=L$ and $N_x=L/N_p)$. Using $L=1024, D/F=10^5, E_1=0.1$ eV, and a step-edge barrier $E_b=0.07$ eV at $T=300$ K, we have carried out multilayer simulations of growth using the reversible model up to a coverage of  $10$ ML. In this case  the parallel efficiency may be written as, 
\begin{equation}
PE = \frac {t_{1p}}{N_p\ t_{av}(N_p)}, \label{eqn7}
\end{equation} 
where $t_{1p}$ is the calculation time for a serial simulation of the $L\times L$ system.  We expect that  in this case the parallel efficiency will decrease more rapidly with increasing $N_p$  than for the case of fixed processor size, since the decreased processor size  leads to increased fluctuations and communications overhead. 
Using the event optimization method, the parallel efficiencies obtained were 28\% ($N_p=4$), 18\% ($N_p=8$) and 9\% ($N_p=16$), respectively, which corresponds to an approximate $N_p^{-1}$ dependence for the parallel efficiency.

\section {Discussion}

 We have carried out parallel kinetic Monte Carlo (KMC) simulations  of  three different simple models of thin-film growth using the synchronous relaxation (SR) algorithm for parallel discrete-event simulation.   
In particular, we have studied the dependence of the parallel efficiency on  the processor size, number of processors, and cycle length $T$, as well as the ratio $D/F$ of  the monomer hopping rate $D$  to the  (per site) deposition rate $F$.  A variety of  techniques for optimizing the parallel efficiency were also  considered.   As expected since the SR algorithm is rigorous, excellent agreement  was found with serial simulations.

Our results indicate that while reasonable parallel efficiencies may be obtained for a small number of processors, due to the requirement of global communications and the existence of fluctuations, the SR algorithm does not scale, i.e. the parallel efficiency decreases logarithmically as the number of processors increases.  
In particular, for the fractal and edge-diffusion models with event optimization, we have found that  the dependence of the parallel efficiency as a function of $N_p$   may be fit to the form $PE = [1 +  c~ (\ln{N_p})^\beta]^{-1}$ where $\beta \simeq 1.5$.  If the communication time is negligible compared to the calculation time, then the parallel efficiency is higher but a similar fit is obtained   with an exponent   close  to $1$, i.e.  $\beta \simeq 1.1$. 
These results   suggest  that while the SR algorithm may be reasonably efficient for a moderate number of processors, for a very large number of processors the parallel efficiency may be unacceptably low. These results are also  in qualitative agreement with the  analysis  presented in Ref.~\onlinecite{Lubachevsky2}   that for parallel Ising spin simulations 
using the SR algorithm with a fixed cycle length, the parallel efficiency should decay as $1/(log  N_p)$ for large $N_p$.

We have also studied in detail the  three main  factors which determine the parallel efficiency in the SR algorithm.  The first is the extra calculation overhead due to  relaxation iterations   which are  needed to correct  boundary events in neighboring  processors.  As expected, the number of relaxation iterations  $I'$ is proportional to the number of boundary events and is also roughly proportional to both   the cycle length $T$ and the range of interaction.   As a result, decreasing the cycle length will decrease the overhead due to relaxation iterations.

 The second main factor determining the parallel efficiency is the  relative (extreme)  fluctuation $\Delta n^e/n_{av}$  in the number of events over all processors in each iteration.  For a  fixed number of processors the relative fluctuation decreases as one over the square-root of the number of events per cycle and is  thus inversely proportional to  the square-root of the product of the processor size and the cycle length.  As a result, decreasing the cycle length $T$ will increase the overhead due to fluctuations.  However,  for a fixed processor size and cycle length  the    relative  fluctuation   also increases logarithmically with the number of processors.  This increase in the relative fluctuation  also leads to an increase in the number of relaxation iterations with increasing $N_p$ as well as    decreased  processor utilization    in each iteration.  As a result the parallel efficiency decreases  as the number of processors increases. 
 
The third factor determining the parallel efficiency is the overhead due to local and global communications.  For the   KMC models we have studied the calculation time per event is   smaller than the latency time due to local communications.  As a result, in our simulations the optimal parallel efficiency was obtained by using a cycle length such that $n_{av} >> 1$ where $n_{av}$ is the average number of events per processor per cycle.   In general, the optimal value of $n_{av}$   may   be determined by balancing the overhead due to relaxation iterations and fluctuations with the overhead due to communications.  Since the global communications time increases logarithmically with the number of processors, the communications overhead also leads to a  decrease in the parallel efficiency with increasing $N_p$.

In order to optimize the parallel efficiency,  we have considered and applied several   techniques.  These include  (i) carrying out several simulations with a different fixed time interval $T = \phi/D$ in order to determine the optimum value of $\phi$, (ii) using direct feedback to dynamically control the cycle length   during a simulation in order to  maximize the ratio of the average number of events per cycle  $n_{av}$ to the measured or estimated execution time, and  (iii) using feedback to dynamically control the cycle length   in order to obtain a pre-determined ``target number" for an auxiliary quantity such as the number of events per cycle or the number of iterations per cycle. While the first  two methods   are the most direct, we have found that in most cases, the third method  results in the highest parallel efficiency. However, since there is no a priori way of knowing the optimal target number in advance, this optimization method must be accompanied by additional simulations.

For the case of negligible communication time, corresponding to simulations in which the calculation time is much longer than the communication time,    the cycle time should  be small in order to minimize the number of additional iterations but not too small  since a very small cycle time will lead to large relative  fluctuations.      
For a processor size $N_x = 256, N_y = 1024$, we found that   $n_{opt} \simeq 6$ leads to  ideal parallel efficiencies which were significantly larger than obtained using event optimization with the communications time taken into account.

We note that in our simulations we have focused primarily on the case of strip decomposition  in order to minimize the communications overhead. However, if the calculation time is significantly larger than the communications time, then for a square system  the  parallel efficiency may be somewhat larger if a square decomposition is used instead.  To illustrate this we consider the decomposition of an $L$ by $L$ system into $N_p$ domains. If the width of the boundary region or range of interaction  is given by $w$,  then for the case of  strip decomposition the area of the boundary region   in each processor  is given by $A_{bdy}  = 2 w L/N_p$. However ignoring corner effects, the area of the boundary region for the square decomposition case is given by $A_{bdy}  = 4 w L/\sqrt{N_p}$.  For $N_p > 4$, the area of the boundary region   is smaller for square decomposition than for strip decomposition. 
Since the number of iterations is roughly proportional to the area of the boundary region, the  calculation overhead due to relaxation iterations will be larger for $N_p > 4$  for the case of strip decomposition.  As a result, we expect 
that for a fixed (square) system size and a large number of processors, and for systems (unlike those studied here)  with a high ratio of (per event) calculation time to communications time,  square decomposition may be significantly more efficient than strip decomposition.

We also note that in our simulations we have used two slightly different definitions for  the  parallel efficiency. 
In the first definition (Eq. \ref{eqn4}),  the parallel execution time was compared with  the serial execution time of a system whose size is the same as a single processor.  In contrast, in the second definition (Eq. \ref{eqn7}) the parallel execution time was directly compared with $1/N_p$ times the serial execution time of a system whose total system size is the same as   in the parallel simulation. If the serial KMC  calculation time per event is independent of   system size, then there should be no difference between the two definitions.  Since in the   models studied here we have used lists for each type of event,  we would expect the serial calculation time per event to be   independent of  system size, and thus the two definitions of parallel efficiency should  be equivalent. To test if this is   the case, we have calculated the serial simulation time per event for the fractal model for $D/F = 10^3$ and $D/F = 10^5$ for a variety of system sizes ranging from $L = 64$ to $L = 2048$.  Somewhat surprisingly, we found that the serial calculation time per event increases slowly with increasing processor size.  In particular,  an increase of approximately 50\%  in  the calculation time per event was obtained when going from a system of size $L = 64$ to  $L = 2048$.    We believe that this is most likely due to  memory or ``cache" effects in our simulations.  This increase in the serial calculation time per event with increasing system size indicates that the  calculated  parallel efficiencies  shown in Fig.~\ref{fig:Fig14} and  Fig.~\ref{fig:Fig15} would  actually be somewhat  larger if the more direct definition of parallel efficiency (Eq. \ref{eqn7}) were used.

We now discuss  some possible improvements of the method  described  here.  As already noted, in our  parallel KMC simulations, lists were used in order to maximize the serial efficiency.  However, for simplicity each additional iteration after the first iteration was restarted at the beginning of the cycle rather than starting with the first new boundary event in each processor.  By using the more efficient method of only redoing events starting with the first new boundary event, we expect a possible  maximum increase in the parallel efficiency of approximately 25\% over the results presented here. 
In addition, it is also possible that by using improved feedback methods, the parallel efficiency may  be somewhat further increased.  For example, by modifying the feedback algorithm  it may be possible   to further improve the direct optimization  method.  It may also be possible to combine all three optimization methods to obtain an improved  parallel efficiency for a given simulation.

Finally, we note that the main reason for the low parallel efficiency for a large number of processors is the global requirement that all processors must be perfectly synchronized. However,  for systems with short-range interactions  it should be possible to at least temporarily relax this synchronization requirement  for processors which are sufficiently  far away  from one another.  Thus, in large systems with a large number of processors it may be possible to increase the parallel efficiency by slightly modifying the SR algorithm by making it somewhat less restrictive.  
In this connection, we have recently developed \cite{yshimsublattice} a semi-rigorous synchronous sublattice algorithm which yields excellent agreement  with serial simulations for all but the smallest processor sizes and in which the asymptotic parallel efficiency is constant  with increasing processor number.  By combining the synchronous sublattice algorithm with the SR algorithm it may be possible to obtain a hybrid algorithm which contains the best features of both   e.g. accuracy and efficiency.

\begin{acknowledgments}
This research was supported by the NSF through Grant No. DMR-0219328. We would also like to acknowledge grants of computer time from the Ohio Supercomputer Center (Grant No. PJS0245) and the Pittsburgh Supercomputer center (Grant No. DMR030007JP).

\end{acknowledgments}


\begin{thebibliography}  {99}

\bibitem{bkl} A. B. Bortz, M.H. Kalos, and J. L. Lebowitz,  J. Comp. Phys. {\bf 17} 10 (1984). 

\bibitem {voter} A. F. Voter, Phys. Rev. B {\bf 34}, 6819 (1986).

\bibitem {maksym} P. A. Maksym, Semicond. Sci. Technol. {\bf 3}, 594 (1988).

\bibitem {fichthorn} K. A. Fichthorn and W. H. Weinberg, J. Chem. Phys. {\bf 95}, 1090 (1991). 


\bibitem {blue} J. L. Blue, I. Beichl, and F. Sullivan,  Phys. Rev. E {\bf 51},  R867 (1995). 

\bibitem{Metropolis} N. C. Metropolis, A. W. Rosenbluth, M. N. Rosenbluth, A. H. Teller, and E. Teller,  J. Chem. Phys. {\bf 21}, 6 (1953). 

 \bibitem{chandy} K. M. Chandy and J. Misra, IEEE Trans. Software Eng. {\bf 5}, 440 (1979); J. Misra, ACM Comput. Surv. {\bf 18}, 39 (1986). 

\bibitem{Lubachevsky1} B. D. Lubachevsky, Complex Systems {\bf 1}, 1099 (1987); J. Comput. Phys. {\bf 75}, 103 (1988).

\bibitem{Korniss1} G. Korniss, Z. Toroczkai, M. A. Novotny, and P. A. Rikvold, Phys. Rev. Lett. {\bf 84}, 1351 (2000). 

\bibitem{Korniss2} G. Korniss, M. A. Novotny, Z. Toroczkai, and P. A. Rikvold, in {\it Computer Simulated Studies in Condensed Matter Physics XIII}, D. P. Landau, S. P. Lewis and H.-B. Schuttler eds. Springer Proceedings in Physics, Vol. 86 (Springer-Verlag, Berlin Heidelberg, 2001).

\bibitem{Korniss3} G. Korniss, M. A. Novotny, H. Guclu, Z. Toroczkai, and P. A. Rikvold, Science {\bf 299}, 677 (2003).

\bibitem {Kornissjcp} G. Korniss, M. A. Novotny, and P. A. Rikvold, J. Comp. Phys. {\bf 153}, 488 (1999). 

\bibitem{Kornisse1} G. Korniss, C. J. White, P. A. Rikvold, and M. A. Novotny, Phys. Rev. E {\bf 63}, 016120 (2001).

\bibitem{Kornisse2} G. Korniss, P. A. Rikvold, and M. A. Novotny, Phys. Rev. E {\bf 66}, 056127 (2002).

\bibitem{yshimnovotny} Y. Shim and J. G. Amar, unpublished. 

\bibitem{yshimsublattice} Y. Shim and J. G. Amar, cond-mat/0406379.

\bibitem{Eick} S. G. Eick, A. G. Greenberg, B. D. Lubachevsky, and A. Weiss, ACM Transactions on Modeling and Computer Simulation {\bf 3}, 287 (1993).

\bibitem {Lubachevsky2} B. D. Lubachevsky and A. Weiss, in {\it Proceedings of the 15th Workshop on Parallel and Distributed Simulation} (PADS'01) IEEE (2001).  See also http://arXiv.org/abs/cs.DC/0405053. 


\bibitem{islandprb} J. G. Amar, F. Family, and P.-M. Lam, Phys. Rev. B {\bf 50}, 8781 (1994). 

\bibitem {ES} G. Ehrlich and F. Hudda,  J. Chem. Phys. {\bf  44}, 1039 (1966); R.L. Schwoebel, J. Appl. Phys. {\bf  40}, 614 (1969).

\bibitem{Venables} J. A. Venables, Philos. Mag. {\bf 27}, 697 (1973). J. A. Venables, G. D. Spiller, and M. Hanbucken, Rep. Prog. Phys. {\bf 47}, 399 (1984).

\bibitem{islandprl} J. G. Amar and F. Family, Phys. Rev. Lett. {\bf 74}, 2066 (1995). 

\bibitem {Zhangdimer} Z.-P. Shi, Z. Zhang, A. K. Swan, and J.  F. Wendelken, Phys. Rev. Lett. {\bf 76}, 4927 (1996). 

\bibitem{Hamilton} J. C. Hamilton, M. R. Sorensen, and A. F. Voter, Phys. Rev. B {\bf 61}, R5125  (2000). 

\bibitem{Jonsson} G. Henkelman and H. Jonsson, Phys. Rev. Lett. {\bf 90}, 116101 (2003). 

\bibitem{tad4} M. R. Sorensen and A. F Voter, J. Chem. Phys. {\bf 112}, 9599 (2000);  A. F. Voter, F. Montalenti, and T. C. Germann, Annu. Rev. Mater. Res. {\bf 32}, 321 (2002). 


\end{thebibliography}
\end{document}